\documentclass[twocolumn,showpacs,preprintnumbers,amsmath,amssymb,prb,superscriptaddress,nobibnotes,nofootinbib]{revtex4}
\usepackage[latin1]{inputenc} \usepackage{subfigure}

\newcommand{\eq}[1]{(\ref{eq:#1})} 
\newcommand{\diffd}[0]{\mathrm{d}}
\newcommand{\iu}[0]{\mathrm{i}} 
\newcommand{\ex}[0]{\mathrm{e}}

\newcommand{\mc}[1]{\mathcal{#1}}
\newcommand{\mb}[1]{\mathbf{#1}}

\newcommand{\mat}[2]{\left(\!\!\begin{array}{#1} #2 \end{array}
    \!\!\right)}

\usepackage{graphicx}
\usepackage{bm}

\begin{document}

\title{Decoherence-induced conductivity in the discrete 1D Anderson model:
\\A novel approach to even-order generalized Lyapunov exponents}

\author{Mat\'ias Zilly} \email{matias.zilly@uni-due.de}
\affiliation{Department of Physics, University of Duisburg-Essen and
CeNIDE, 47048 Duisburg, Germany} 
\author{Orsolya Ujs\'aghy}
\affiliation{Department of Theoretical Physics and Condensed Matter
Research Group of the Hungarian Academy of Sciences,\\ Budapest
University of Technology and Economics, Budafoki \'ut 8., H-1521
Budapest, Hungary} 
\author{Marko Woelki} \affiliation{Department of Physics, Saarland University, 66123 Saarbr\"ucken,
Germany} 
\author{Dietrich E.\ Wolf}
\affiliation{Department of
Physics, University of Duisburg-Essen and CeNIDE, 47048 Duisburg,
Germany}

\date{\today}

\begin{abstract} A recently proposed statistical model for the effects
  of decoherence on electron transport manifests a decoherence-driven
  transition from quantum-coherent localized to ohmic behavior when
  applied to the one-dimensional Anderson model. Here we derive the
  resistivity in the ohmic case and show that the transition to
  localized behavior occurs when the coherence length surpasses a
  value which only depends on the second-order generalized Lyapunov
  exponent $\xi^{-1}$. We determine the exact value of $\xi^{-1}$ of
  an infinite system for arbitrary uncorrelated disorder and electron
  energy. Likewise all higher even-order generalized Lyapunov
  exponents can be calculated, as exemplified for fourth order. An
  approximation for the localization length (inverse standard Lyapunov
  exponent) is presented, by assuming a log-normal limiting
  distribution for the dimensionless conductance $T$. This
  approximation works well in the limit of weak disorder, with the
  exception of the band edges and the band center.
\end{abstract}

\pacs{72.10.-d,72.15.Rn,71.23.An}
\maketitle

\section{Introduction}
\label{sec:intro}

Anderson \cite{anderson58} was the first to show that electronic
eigenstates in disordered media can be localized, if the disorder
exceeds a certain threshold. In the infinite, discrete one-dimensional (1D)
Anderson model 
\begin{align}
  \label{eq:H_lin_chain} H = \sum_{i} \epsilon_i | i \rangle \langle
i| + t \sum_{i}[| i \rangle \langle i+1 | + \text{h.c.} ]
\end{align} 
without correlations, where the onsite energies $\epsilon_i$ are
independently distributed according to a probability density
$w(\epsilon)$ (with mean value 0, variance $\sigma^2$, third and
fourth moments $\nu^3$ and $\kappa^4$), all eigenstates are localized
for $\sigma^2>0$, i.e.~the eigenfunctions decay exponentially in
space. To simplify the notation we take $t=1$ as energy unit and the
lattice spacing $a=1$ as length unit in the
following.

Even fifty years after its foundations, the physics of Anderson
localization is a very active field of research, stimulated by the
recent observations of localization of light and cold atoms,
cf.~Refs.\ \onlinecite{lagendijk09,evers08,modugno10} and references
therein. 

The ``absence of diffusion'' due to disorder leads to an exponential
increase of the electrical resistance of the localized system, if the
electrons behave coherently.\cite{anderson58,berezinskii74} Ohmic
behavior, a resistance increase proportional to the linear extension
of the device, can be achieved in spite of localization, if sufficient
decoherence is included.\cite{datta95} In recent publications we have
demonstrated this using a statistical model for the effects of
decoherence.\cite{zilly09,zilly10,zilly10PhD} Also the experimental
observation of sequence-dependent conductance in DNA molecules (ohmic
vs.\ exponential) \cite{xu04} could be explained.\cite{zilly10-2}

In this article we examine the decoherence-induced conductivity in the
Anderson model subjected to the phenomenological model for
decoherence. 

Decoherence-induced conductivity in disordered systems
was already subject of various papers.\cite{gogolin76,hey95,maslov09}
Here we examine how much decoherence is necessary in order to obtain 
finite conductivity and conclude that whenever the
coherence length $L_\phi$ fulfills
\begin{align}
  \label{eq:ohm_limit}
  L_\phi < \frac{1}{1-\exp(-\xi^{-1})},
\end{align}
ohmic conductance is reached. Here $\xi^{-1}$ is the second-order
generalized Lyapunov exponent
(GLE),\cite{paladin87,zillmer03,gurevich11} for which we report exact
results for arbitrary diagonal disorder and electron energy. All
higher even-order GLEs, e.g.~the fourth-order GLE $\chi^{-1}$, can be
calculated analytically by the same method as $\xi^{-1}$. The
even-order GLEs have also been determined for nonlinear
oscillators.\cite{mallick02}

Assuming a Gaussian limiting distribution for the logarithm of the
dimensionless conductance $\ln T$, we present an approximation in which the
localization length $\lambda$ (the inverse standard Lyapunov
exponent LE) is determined by $\xi$ and $\chi$. This
approximation works well in the limit of weak disorder
with the exception of the band edges and the band center.

The paper is organized as follows. Sect.~\ref{sec:model} explains the
coherent transport formalism and the phenomenological statistical
model by which we include decoherence effects. In
Sect.~\ref{sec:resistivity} we derive the resistivity and the
limitations under which finite conductivity is
found. Sect.~\ref{sec:xi} introduces the GLEs and derives
$\xi^{-1}$. Using the GLEs, in Sect.~\ref{sec:loclen} we perform an
approximate calculation of the localization length. Sect.~\ref{sec:conclusions}
discusses the results and conclusions. The Appendices contain details
of the calculations.

\section{Model}
\label{sec:model}

\begin{figure}[htbp] \centering
\subfigure[]{\label{fig:IntroDecBonds}\includegraphics[scale=0.5]{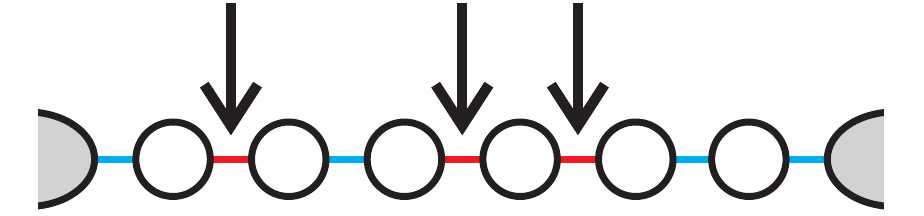}}
\\
\subfigure[]{\label{fig:ResultingChainsDecBonds}\includegraphics[scale=0.5]{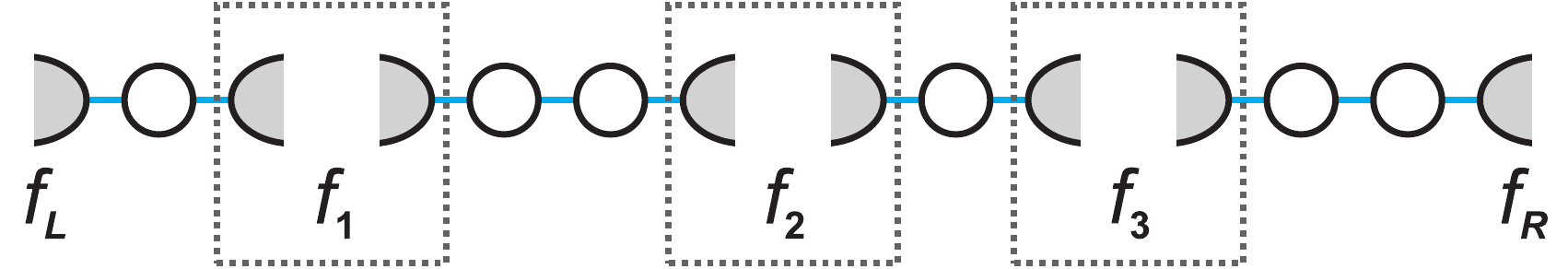}}
  \caption{(Color online) A coherent chain of $N=6$ sites coupled to
wide-band contacts.  \subref{fig:IntroDecBonds} Three decoherence
bonds introduced at random according to a probability $p$ form a
decoherence configuration.  \subref{fig:ResultingChainsDecBonds} The
resulting equivalent set of coherent chains. The $f_k$ denote the
electron energy distribution functions at the contacts and decoherence
bonds.}
  \label{fig:Decoherence_Models}
\end{figure}

The idea of the statistical model for the effects of decoherence on
electron transport is the following.\cite{zilly09} We describe the
electrons in a single-electron picture. Decoherence is modeled by
stochastic events: The electron, which in general is considered
coherently, is extracted and re-injected at decoherence events, which
take place at random locations in space. Thereby it loses its
coherence completely. The quantity of interest, e.g.\ the conductance
\cite{zilly09,zilly10PhD,zilly10-2} or resistance
\cite{zilly10,zilly10PhD} is calculated for a given \emph{decoherence
configuration}, a division of the system into coherent and decoherence
regions. Afterwards the quantity of interest is averaged over an
ensemble of decoherence configurations. Here we apply our statistical
decoherence approach to the Anderson model in the following way,
cf.\ Fig.~\ref{fig:Decoherence_Models}, which is a slight modification
of the original.

According to the probability $p$, the bonds of the Anderson
Hamiltonian $H$ are replaced by \emph{decoherence bonds}, virtual
contacts in the wide-band limit which cause a purely imaginary
self-energy $-\iu \eta$ to the neighboring regions of the system,
forming a decoherence configuration. Thus $p$ and $\eta$ are the two
parameters by which we describe phenomenologically the effect of
decoherence.

A decoherence bond represents the environmental decohering effect, or
a decoherent escape into the rest of the electronic system. An
electron is extracted out of the system at the decoherence bond (the
bond represents a reservoir) and re-injected with the same energy (no
inelastic effects are considered) at the same bond involving a
complete loss of phase coherence.

The Green's function of a \emph{coherent subsystem} which is delimited
by two decoherence bonds and contains the sites
$|1\rangle$--$|j\rangle$ reads
 \begin{align}
\begin{split}
  \label{eq:G_coherent_channel} G &\equiv \left[E-H+ \iu \eta
(|1\rangle\langle 1| + |j\rangle\langle j|)\right]^{-1},
\end{split}
\end{align} 
where $H$ is the Hamiltonian restricted to the sites
\emph{between} the two decoherence bonds, and $E$ is the total electron
energy. Due to the simple structure of the self-energies, the
transmission through the subsystem according to the Nonequilibrium
Green's function method (NEGF) \cite{datta95} reads
\begin{align}
  \label{eq:Transmission} T_j = 4 \eta^2 | \langle 1 | G | j \rangle
|^2
\end{align} and can be evaluated recursively because of the
tridiagonal structure of $G^{-1}$:
\begin{align} 
\label{eq:1_T_j_r}
\frac{1}{T_j} = \frac{|r_j-\eta^2 s_{j-1}+\iu \eta (
r_{j-1}+ s_j)|^2 }{4\eta^2},
\end{align} 
with the polynomials
\begin{equation}
\begin{aligned}
  \label{eq:iteration_r_j} r_j&=(E-\epsilon_j)r_{j-1}-r_{j-2}, &
s_j&=(E-\epsilon_j)s_{j-1}-s_{j-2},\\ r_0&=1, & s_1&=1, \\ r_{-1}&=0,
& s_0&=0.
\end{aligned}
\end{equation}
For the derivation of Eqs.\ \eq{1_T_j_r} and \eq{iteration_r_j}, see
the Appendix~\ref{sec:Green_iter}.

Although the dimensionless conductance $T$ is often calculated
without making reference to a coupling to external
leads,\cite{kramer93} the results for large coherent systems which we
discuss below do not depend on the model parameter $\eta$, as it
becomes an irrelevant boundary condition (cf.\ Appendix
\ref{sec:without_leads}).

Under the condition of a complete loss of phase coherence without
energy relaxation at the decoherence bonds, we have shown in
Ref.\ \onlinecite{zilly09}, that the resistance of a decoherence configuration at
infinitesimal bias is the sum of the individual subsystem resistances,
\begin{align}
  \label{eq:R_config} R = \frac{h}{2e^2} \sum_k \frac{1}{T_k},
\end{align} 
where $k$ enumerates the coherent subsystems, and
$\tfrac{2e^2}{h}=1$ is the quantum of conductance assuming spin
degeneracy, which we take as our conductance unit. To obtain the resistance as a
system parameter, $R$ is to be averaged over all decoherence
configurations.

Now consider the thermodynamic limit of an infinite system. As the
locations of the decoherence bonds are uncorrelated, the relative
frequency of the subsystem size $j$ is $pq^{j-1}$, where
\begin{align}
   \label{eq:def_q} 
	q\equiv 1-p
\end{align} 
is the probability of a regular bond. On average, the
subsystem size is 
\begin{align}
\label{eq:L_phi}
L_\phi=\frac{1}{p},
\end{align} 
defining the coherence length in our model. In the infinite system,
subsystems of size $j$ appear infinitely many times, contributing on
average $\langle 1/T_j \rangle$ to the resistance. Here,
$\langle\cdot\rangle$ denotes the average over the disorder.

\section{Resistivity of the system}
\label{sec:resistivity}

We define the resistivity of the system as the resistance per length in the limit of an infinite system:
\begin{align}
  \label{eq:resistivity} \rho = p^2 \sum\limits_{j=1}^\infty
q^{j-1}\left \langle\!  \frac{1}{T_{j}}\!\right \rangle.
\end{align} 
Naturally this definition is restricted to those values
$q<q^\ast$ for which the series converges. Then the system
behaves ohmically. For $q>q^\ast$, localization dominates, and the
resistance increases exponentially with the system size. This is the
decoherence induced conductivity: As the decoherence density $p$
increases over the critical value $p^\ast=1-q^\ast$, there is a
transition from localized to ohmic behavior.

The disorder average $\langle 1/T_j \rangle$ is a polynomial
of order $2j$
\begin{align}
  \label{eq:def_T_j} 
  \begin{split}
    \left\langle\! \frac{1}{T_j}\!\right\rangle
    &\equiv \int\limits_{-\infty}^{\infty}\!\!\! \diffd \epsilon_1 \dots
    \diffd \epsilon_j \frac{1}{T_j} \prod\limits_{i=1}^j w(\epsilon_i)\\
     &= \frac{1}{2} + \frac{1}{4\eta^2}
    R_j + \frac{1}{2} R_{j-1} + \frac{\eta^2}{4}
    R_{j-2},
  \end{split}
\intertext{where we define}
  \label{eq:defR_j} 
  R_j &\equiv
  \int\limits_{-\infty}^{\infty}\!\!\! \diffd \epsilon_1 \dots \diffd
  \epsilon_j r_j^2 \prod\limits_{i=1}^j w(\epsilon_i).
\intertext{Using Eq.~\eq{iteration_r_j} we find a recursion for the
$R_j$:}
  \begin{split} 
    R_j&=\int\limits_{-\infty}^{\infty}\!\!\!
    \diffd \epsilon_1 \dots \diffd \epsilon_j \left[ (E^2-2E\epsilon_j+
      \epsilon_j^2)r_{j-1}^2 \right.\\[-0.4cm] & \qquad\qquad\left.+
      2(\epsilon_{j}-E)r_{j-1}r_{j-2} + r_{j-2}^2 \right]
    \prod\limits_{i=1}^j w(\epsilon_i)
  \end{split}\nonumber\\
  \label{eq:R_j_recursion}  
    &=\left(E^2+\sigma^2\right)
    R_{j-1} -2 E S_{j-1} + R_{j-2}, 
\intertext{where}
  \label{eq:def_G_j} S_j &\equiv
\int\limits_{-\infty}^{\infty}\!\!\! \diffd \epsilon_1 \dots \diffd
\epsilon_j r_jr_{j-1}\prod\limits_{i=1}^j w(\epsilon_i)
\intertext{itself fulfills the recursion}
    S_j &= \int\limits_{-\infty}^{\infty}\!\!\!
    \diffd\epsilon_1 \dots \diffd \epsilon_j \left[(E-\epsilon_j)
      r_{j-1}^2 - r_{j-1}r_{j-2}\right] \prod\limits_{i=1}^j w(\epsilon_i)
\nonumber    \\
  \label{eq:G_j_recursion}
    &= E R_{j-1}- S_{j-1}.
\end{align} 
The initial conditions $R_{-1}=0$,
$S_0=0$ and $R_0=1$ allow to determine the
disorder-averaged resistance $\langle 1/T_j \rangle$ of a
subsystem of size $j$ via Eqs.\
\eq{R_j_recursion}--\eq{G_j_recursion}.

Using the recursions in Eqs.~\eq{R_j_recursion} and
\eq{G_j_recursion} we calculate the generating function of the
$R_j$:
  \begin{align}
  \label{eq:def_F}
  \begin{split}
 \mc{R}(z)&\equiv
\sum\limits_{j=1}^\infty R_j z^{j-1}\\
   &=\frac{1}{N_1} \left(-z^2+\left(E^2-\sigma^2-1\right) z -
    E^2-\sigma^2\right),
  \end{split}
\end{align}
where
\begin{align}
\label{eq:def_N_1}
N_1&\equiv z^3+\left(1-E^2+\sigma^2\right) z^2 +
    \left(E^2+\sigma^2-1\right) z -1.
\end{align}

With Eq.~\eq{def_T_j} we evaluate the resistivity $\rho$ [Eq.\
\eq{resistivity}] in terms of the generating function
$\mc{R}(z=q)$ arriving at the closed form valid for $q<q^\ast$
\begin{align}
  \label{eq:rho_result}
  \begin{split} \rho&= \frac{p}{2}
+\left[p^2\left(\frac{1}{2}+\frac{1}{4\eta^2}\left(\sigma^2+E^2\right)\right)\right.
\\ &\left.\qquad+ p^2q \left(\frac{1}{4\eta^2} + \frac{\eta^2}{4} -
\frac{2E^2}{4\eta^2(1+q)} \right) \right] \\ &\times
\left[1-\left(\sigma^2+E^2\right)q - q^2 + \frac{2E^2
q^2}{1+q}\right]^{-1}\!\!\!\!\!\!\!\!.
  \end{split}
\end{align}

\begin{figure}[htbp] \centering
  \includegraphics[width=0.45\textwidth]{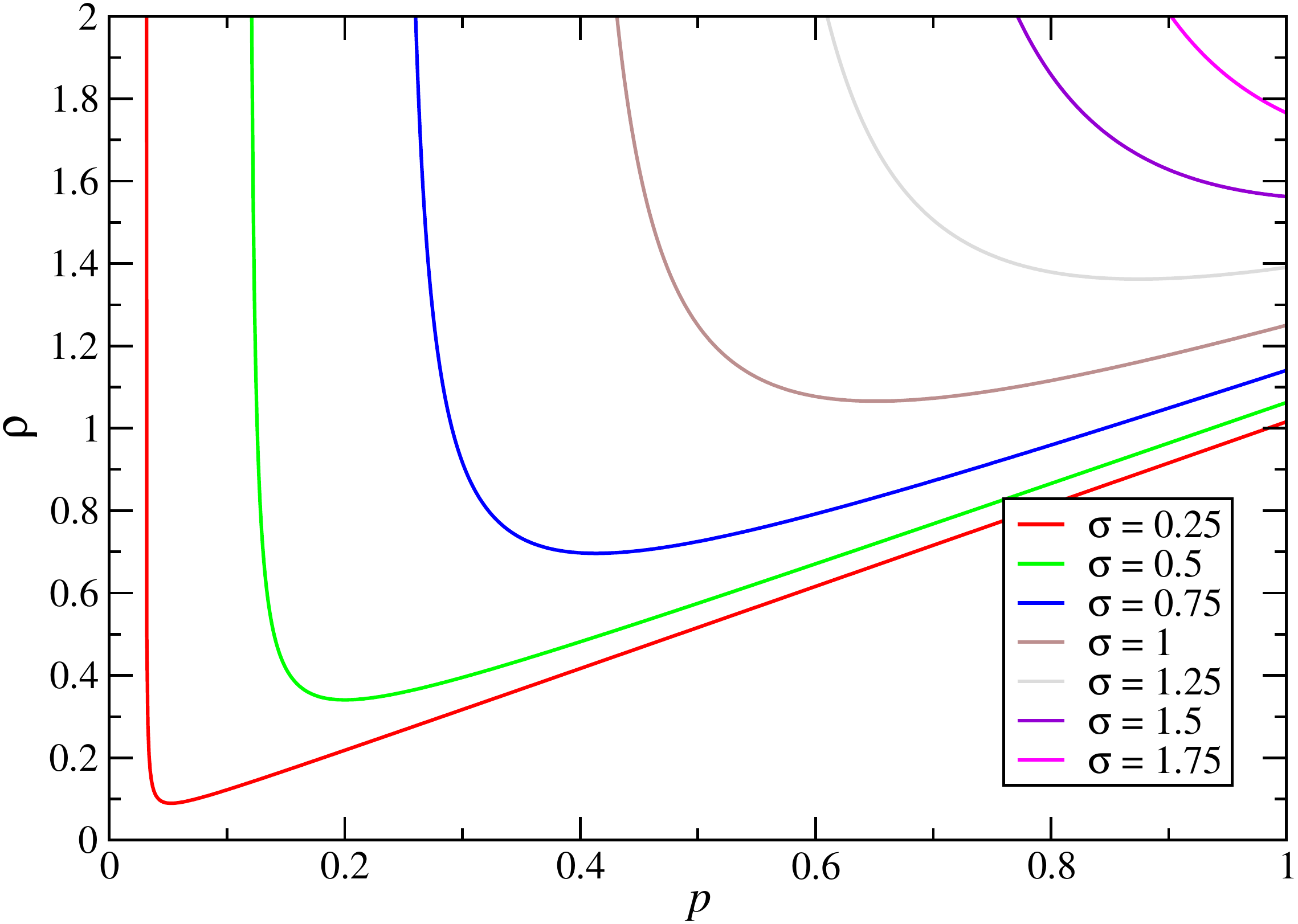}
  \caption{(Color online) The resistivity $\rho$ as a function of the
    density of decoherence events $p$ (=$1/L_\phi$, the inverse
    coherence length) for various disorder strengths
    $\sigma$. For $p<p^\ast$ ($p^\ast$ are the poles in this graph) no finite
    conductivity exists. Parameters: Electron energy $E=0$, broadening $\eta=1$.}
  \label{fig:rho_of_p_diff_W}
\end{figure}

Fig.~\ref{fig:rho_of_p_diff_W} displays the resistivity $\rho$ as a
function of the decoherence density $p$.

The radius of convergence $q^\ast$ is given by the
singularity of $\rho$ which is nearest to the origin. I.e.\ $q^\ast$
is the smallest of the roots of $N_1$, which fulfill the relation
\begin{align}
  \label{eq:ellipse} g(q)\equiv\frac{q}{1-q^2}\sigma^2 +
\frac{q}{(1+q)^2}E^2 =1.
\end{align} 
We see that the singularities lie on an ellipse in the
$\sigma$-$E$ plane, the semiaxes of which are determined by $q$.

Both summands of $g(q)$ are strictly increasing for $q\in[0,1)$,
furthermore $g(0)=0$ and $\lim_{q\rightarrow 1}g(q)=\infty$. Thus Eq.\
\eq{ellipse} obviously has a single, real solution $q^\ast
\in(0,1)$. The two other solutions are complex and fulfill $|q|>1$,
hence they have no physical meaning. This guarantees that for any
disorder and any electron energy there is a critical decoherence
density $p^\ast$ above which ohmic behavior is achieved.

The solutions $q_k$ of Eq.~\eq{ellipse} 
read
\begin{align}
q_k&= \frac{1}{3} (E^2-\sigma^2-1)\nonumber\\
 &+ \frac{1}{3} 2^{\frac{1}{3}}\ex^{\iu \frac{2}{3}k\pi} f_1(E,\sigma)
 \left( f_2(E,\sigma)+\sqrt{27
     f_3(E,\sigma)}\right)^{-\frac{1}{3}}\nonumber\\ 
\label{eq:q_k}
  &-\frac{1}{3}2^{-\frac{1}{3}}\ex^{-\iu \frac{2}{3}k\pi}\left(
      f_2(E,\sigma)+\sqrt{27 f_3(E,\sigma)}\right)^{\frac{1}{3}}
\end{align}
where the functions
\begin{align}
  f_1(E,\sigma)& = -4+5E^2 -E^4+\sigma^2+2E^2\sigma^2-\sigma^4
  \nonumber \\
  f_2(E,\sigma)& = -16-24 E^2+15 E^4-2E^6+6\sigma^2-12E^2\sigma^2
  \nonumber \\ & +6
  E^4\sigma^2-3\sigma^4 -6E^2\sigma^4+2\sigma^6 \nonumber\\
  f_3(E,\sigma)&= 64 E^2-48E^4+12 E^6-E^8-4\sigma^4 + 20 E^2\sigma^4 \nonumber\\ &+
  2 E^4\sigma^4 -\sigma^8\nonumber
\end{align}
are polynomials and $k=0,1,2$. Note that by
$\sqrt{\cdot}$ and $(\cdot)^{\frac{1}{3}}$ we denote the solution of
$z^2 = \cdot$ and $z^3 = \cdot$ with the smallest complex argument
$\arg(z) \in[0,2\pi)$, respectively.
Using \eq{q_k} we can write
\begin{align}
  \label{eq:def_q^ast}
  q^\ast=\min_k\{|q_k|\}=1-p^\ast.
\end{align}

\begin{figure}[htbp] \centering
  \includegraphics[width=0.45\textwidth]{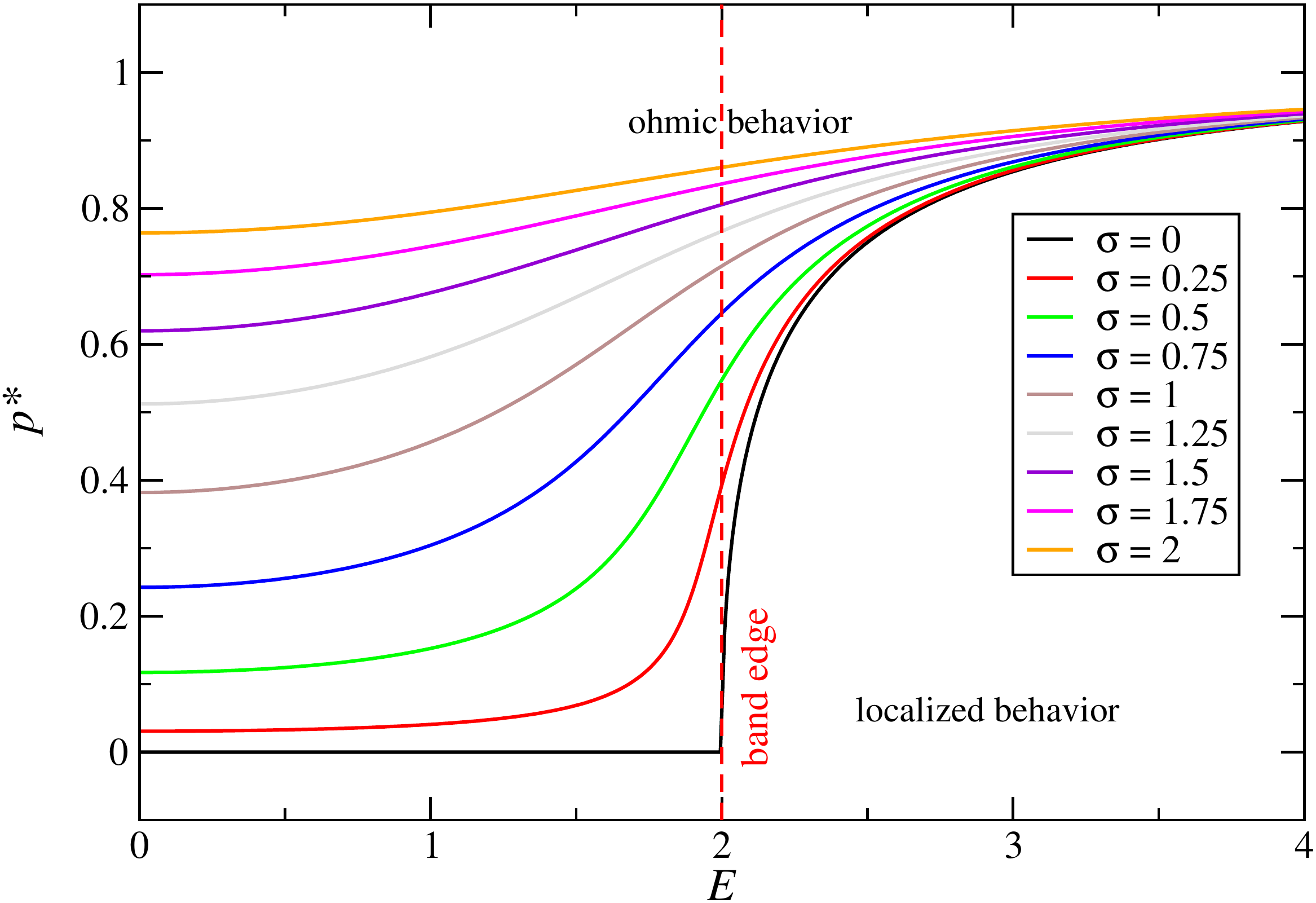}
  \caption{(Color online) The critical decoherence density $p^\ast$
for various values of $\sigma$ as a function of the electron energy
$E$. For $p>p^\ast$, ohmic behavior is achieved, whereas for
$p<p^\ast$ the resistance increases exponentially with the system
size. In the limit $\sigma\rightarrow 0$ we observe a kink at $E=2$,
the band edge of the system without disorder.  }
  \label{fig:p_ast_of_mu_diff_W}
\end{figure}

The dependence of $p^\ast$ on $E$ is displayed in Fig.\
\ref{fig:p_ast_of_mu_diff_W}.

For the interpretation of Fig.~\ref{fig:p_ast_of_mu_diff_W} note that
the range $E\in[-2,2]$ is the energy band of the system without
disorder. Thus only for $|E|\leq 2$ there are eigenstates, which in
this case are extended states. Electrons with $|E|>2$ effectively have
to tunnel through the system. This only leads to a finite resistivity
if the tunneling length is reduced by increasing the decoherence
parameter $p$. The physical picture is similar for the case with
disorder. Now there are no extended states, and electron transmission
is exponentially suppressed for any $E$. Thus $p^\ast>0$ for all values of $E$.

\section{The generalized Lyapunov exponent {\boldmath $\xi^{-1}$}}
\label{sec:xi}

The generalized Lyapunov exponents (GLEs) are defined as
\cite{paladin87,zillmer03,gurevich11}
\begin{align}
L(r) \equiv \frac{1}{r}\lim \limits_{j\rightarrow \infty} \frac{1}{j} \ln \left\langle T_j^{-\frac{r}{2}} \right\rangle.
\end{align}
E.g.\ the second-order GLE $\xi^{-1}\equiv 2L(r=2)$ reads 
\begin{align}
\xi^{-1}= \lim\limits_{j\rightarrow \infty} \frac{1}{j}
\ln \left\langle\! \frac{1}{T_j} \!\right\rangle. 
\end{align}

Since, according to Eq.~\eq{resistivity}, $\rho/p^2$ is the generating
function of the $\langle 1/T_j \rangle$, its pole $q^\ast$
determines\cite{wilf94} the asymptotic behavior of $\langle
1/T_j\rangle$, and hence also $\xi^{-1}$:
\begin{align}
\label{eq:calc_xi}
\lim \limits_{j\rightarrow \infty} \left\langle\!
  \frac{1}{T_j}\!\right\rangle \sim \left(\frac{1}{q^\ast} \right)^j
\quad \Rightarrow \quad \xi^{-1}=-\ln(q^\ast).
\end{align}
Therefore $\xi^{-1}$ can be determined exactly for any set of
parameters $(\sigma,E)$ using Eqs.~\eq{q_k}--\eq{def_q^ast} and
\eq{calc_xi}.

Note that apart from the energy, $\xi^{-1}$ only depends on the second
moment $\sigma^2$ of the distribution of the onsite energies. Higher
moments do not enter.

\begin{figure}[htbp] 
  \centering
  \includegraphics[width=0.45\textwidth]{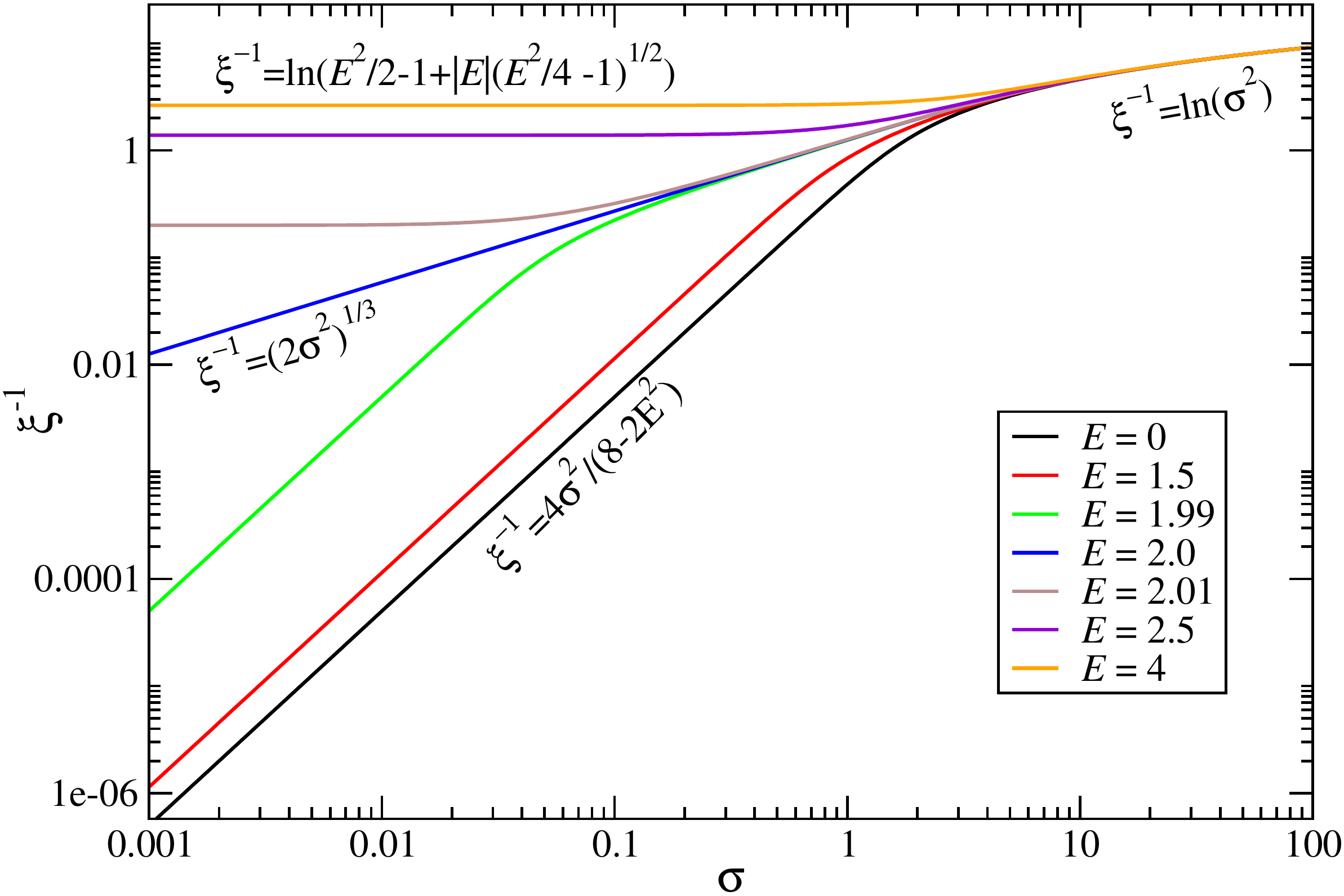}
  \caption{(Color online) Dependence of $\xi^{-1}$ on $\sigma$ for
    different values of $E$.}
  \label{fig:xi_of_sigma}
\end{figure}

Fig.\ \ref{fig:xi_of_sigma} displays the dependence of $\xi^{-1}$ on
the disorder strength $\sigma$ for various energies $E$.

We observe that, depending on $E$ and the disorder strength, $\xi^{-1}$
assumes different functional dependencies on $\sigma$. In the weak
disorder limit $\sigma \rightarrow 0$, $\xi^{-1}$ approximates
\begin{align}
  \label{eq:xi_weak}
  \xi^{-1} &=
  \begin{cases}
    \frac{4\sigma^2}{8-2 E^2}, & |E|<2,\\
 \left(2\sigma^2\right)^{\frac{1}{3}},& |E|=2,
 \vspace{0.1cm}\\
 - \ln \!\left(\!1-\frac{E^2}{2}+\frac{|E|}{2} \sqrt{E^2-4}\!\right),
 & |E|>2 ,
  \end{cases}
\end{align}
and for strong disorder $\sigma\rightarrow\infty$
\begin{align}
  \label{eq:xi_strong} 
  \xi^{-1} =\ln(\sigma^2),
\end{align}
as can be seen easily by inserting $z=\exp(-\xi^{-1})$ into
Eq.~\eq{def_N_1}. 

Outside the band $|E|>2$, $\xi^{-1}$ is finite also
for $\sigma=0$. This is why in Fig.~\ref{fig:xi_of_sigma} and
Eq.~\eq{xi_weak} no $\sigma$-dependence is given outside the band for
weak disorder.

Using $\xi^{-1}$, we can restate the condition $q<q^\ast$ for a finite
conductivity as:
\begin{align}
  \label{eq:L_phi_of_xi}
  L_\phi < \frac{1}{1-\exp(-\xi^{-1})},
\end{align}
which for weak disorder inside the band becomes
\begin{align}
  L_\phi < \xi.
\end{align}

As $\xi^{-1}$ is a property of the infinite coherent system, it is
determined purely by $E$ and $\sigma^2$ and does not depend on the
decoherence parameters $p$ and $\eta$. Yet via Eq.~\eq{L_phi_of_xi},
the second-order GLE $\xi^{-1}$
determines the maximum coherence length for which a finite
conductivity can be reached.

\section{Generalized Lyapunov exponents and localization length}
\label{sec:loclen}

The localization length $\lambda$ is defined as the typical length
scale on which the eigenstates of a disordered system
decay.\cite{thouless79} Its inverse (the standard Lyapunov exponent LE) is related
to the dimensionless conductance of a coherent system via
\begin{align}
  \label{eq:def_lambda}
  \lambda^{-1} = - \lim\limits_{j\rightarrow \infty} \frac{1}{2j}\left\langle \ln
    T_j \right \rangle.
\end{align}

It is easy to see that the localization length and the GLEs are
connected by
\begin{align}
  \label{eq:lambda_GLE}
\lambda^{-1} = \lim\limits_{r\rightarrow 0} L(r).
\end{align}
With our method we can calculate $L(r)$ only for even $r\neq 0$, so
that the limit $r\rightarrow 0$, Eq.~\eq{lambda_GLE}, is not
obvious. It remains to be investigated in the future.

We can, however, use our results to calculate $\lambda$
approximately and gain new insight into the assumptions of single
parameter scaling.

One must distinguish two assumptions, which together define what is
known as single parameter scaling (SPS) in localization
theory.\cite{anderson80,kramer93,shapiro87} 

Assumption A is that the variable $u_j = -\ln T_j$ obeys the central
limit theorem and approaches a Gaussian distribution as $j\rightarrow
\infty$.

Second, assumption B states that a single
parameter suffices to characterize the Gaussian distribution, in other
words, the mean $C_1$ and the variance $C_2$ are proportional to each
other and that the proportionality factor is 2,
\begin{align}
  \label{eq:C2_C1_SPS}
C_2/C_1=2 \quad \text{for SPS.}
\end{align}

Assumptions A and B were proven to hold for the continuous Anderson
model for uncorrelated weak disorder, provided complete phase
randomization takes place between scattering events.
\cite{anderson80,shapiro87}

For the discrete Anderson model, however, already the weaker
assumption A was shown to be wrong at the band edges and the band
center.\cite{schomerus03} Nevertheless, for weak disorder the correct power law
dependencies of $\lambda^{-1}$ on $\sigma$ are reproduced, if
assumption A is applied, see Eqs.\ \eq{lambda_band_center} and
\eq{lambda_band_edge}. Only the prefactors are wrong. For $E$-values
in between the band center and the band edges, assumption A even leads to
the correct $\lambda^{-1}$ in leading order of $\sigma$, see Eq.~\eq{lambda_band}.

In the following we adopt assumption A as an approximation and
investigate how strongly assumption B is violated. We find strong
violation of assumption B at the band edges and at the band
center. However, if for $E$-values in between and weak
disorder, one makes the assumption A, then assumption B is 
automatically fulfilled. 

The assumed Gaussian limiting distribution of the variable $u_j=-\ln
T_j$ is characterized by its mean $C_1$ and variance $C_2$, where
\begin{align}
  \label{eq:C_1_lambda}
\lim\limits_{j\rightarrow\infty} \frac{C_1}{j}=2 \lambda^{-1}.  
\end{align}
This means that the probability $P(u_j)$ to find $u_j\in [u,u+\diffd u]$ fulfills, for a finite
system of length $j$, 
\begin{align}
  \label{eq:central_limit}
   \lim\limits_{j\rightarrow\infty} \left(P(u_j)\diffd u_j -
     \frac{1}{\sqrt{2\pi C_2}}
   \ex^{- \frac{( u_j-C_1 )^2 }{ 2C_2} } \diffd u_j \right)=0,
\end{align}
which allows to relate
\begin{align}
\label{eq:gls_xi_chi_1}
 \begin{split} 
 \xi^{-1} &=\lim \limits_{j\rightarrow \infty} \frac{1}{j} \ln \overline{
    \ex^{u_j}} = \lim \limits_{j\rightarrow \infty} \frac{1}{j} \ln\left[
    \ex^{C_1+C_2/2}\right]\\
  & = \lim \limits_{j\rightarrow\infty} \frac{1}{j}\left(C_1+C_2/2\right),
\end{split}
\end{align}
where $\overline{(\cdot)}$ denotes the average $\int \diffd u_j P(u_j)
(\cdot)$.

Analogously to $\xi^{-1}$, cf.~Eq.~\eq{calc_xi}, we can also calculate the fourth-order
GLE $\chi^{-1}= 4L(4)$ (and similarly, all higher even-order
GLEs),
\begin{align}
  \label{eq:chi_root}
  \chi^{-1} &\equiv \lim \limits_{j\rightarrow \infty} \frac{1}{j} \ln
  \left\langle \! \frac{1}{T_j^2} \! \right\rangle= -\ln(z^\ast)
\end{align}
by determining $z^\ast$, the pole of the generating
function $\mc{T}_{-2}(z)$ of $\langle 1/T_j^2 \rangle$ with the smallest
absolute value.

The definition and derivation of $\mc{T}_{-2}(z)$ can be
found in the Appendix \ref{sec:appendix}. Its denominator reads
\begin{align}
  \label{eq:denominator_S}
N_1 \times N_2 \times (z-1),
\end{align}
where $N_1$ is given in Eq.~\eq{def_N_1} and $N_2$ is a fifth-order
polynomial in $z$ given in Eq.~\eq{Nenner_R}.

Note that, as $1/T_j^2$ contains up to fourth powers of the
polynomials $r_j$ and $s_j$, the roots of $N_2$, and therefore $\chi$
itself, depend also on the third and fourth moments of the probability
distribution $w(\epsilon)$ of the onsite energies. At the same time,
$w(\epsilon)$ only enters through its moments into $\chi$, therefore
any distributions with the same second, third and fourth moments has
the same $\chi$, independently of the exact shape of the
distribution. Analogously $L(6)$ depends on the first six moments of
$w(\epsilon)$, etc.

As $N_2$ is a fifth-order polynomial,
the value of $\chi$ can be calculated within numerical
accuracy. Only for special cases (e.g.\ for $E=0$, see below in Eq.~\eq{chi_E_0}), $\chi$ can be represented in a
closed form. 

Like $\xi^{-1}$ we can relate $\chi^{-1}$ to the cumulants $C_2$ and
$C_1$ under the assumption A:
\begin{align}
  \label{eq:gls_xi_chi_2}
\begin{split}
\chi^{-1} &=\lim \limits_{j\rightarrow \infty} \frac{1}{j} \ln  \overline{
    \ex^{2u_j} } \\
  &=\lim\limits_{j\rightarrow \infty} \frac{1}{j}\left(2C_1+2C_2\right). 
\end{split}
\end{align}
This equation together with Eqs.~\eq{C_1_lambda} and \eq{gls_xi_chi_1}
leads to the localization length
\begin{align}
\label{eq:lambda_final}
  \lambda^{-1} &= \xi^{-1} - \frac{1}{4} \chi^{-1} \quad \text{under
    assumption A.}
\end{align}

\begin{figure}[htbp] 
  \centering
  \includegraphics[width=0.45\textwidth]{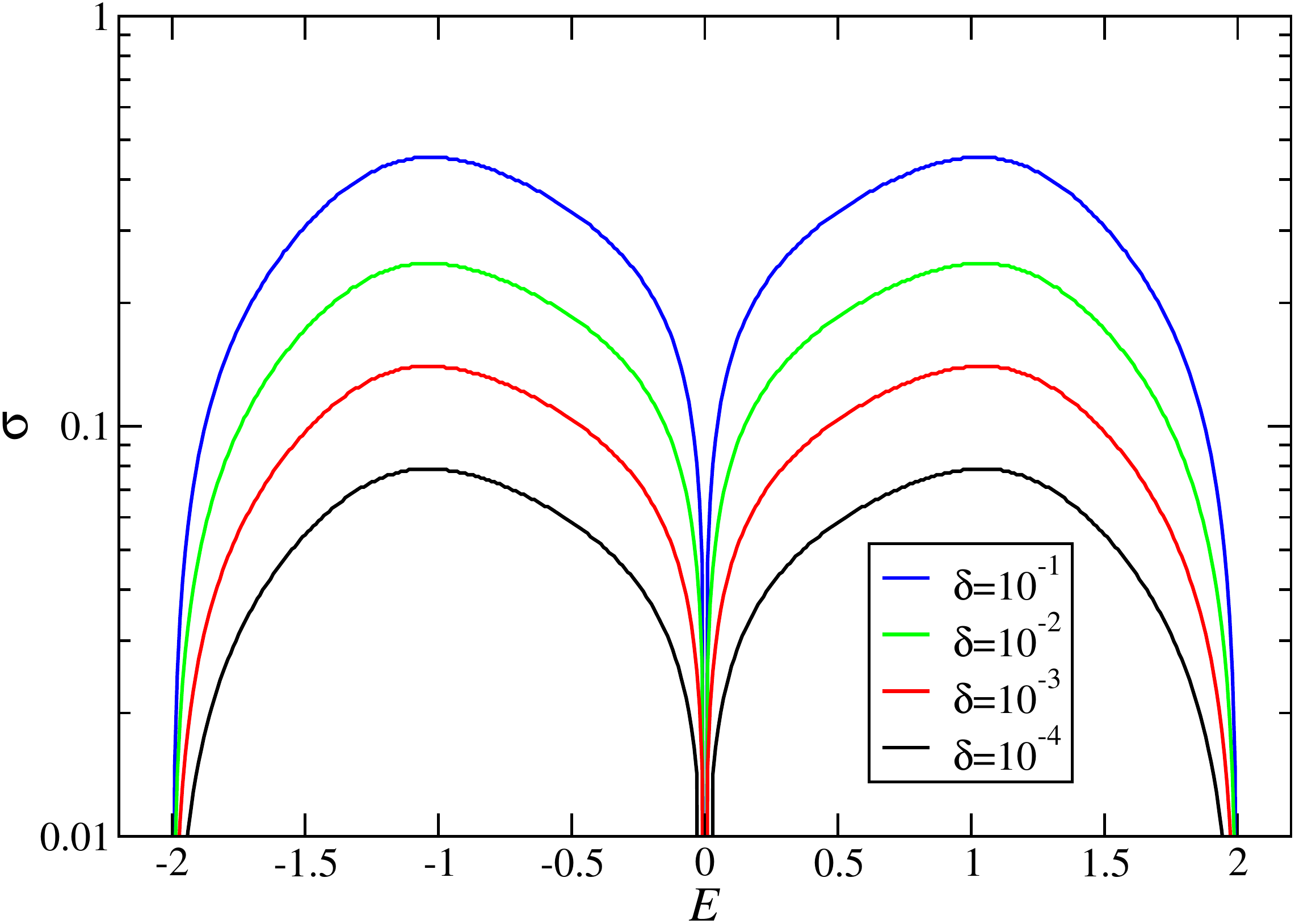}
  \caption{(Color online) Contour lines for $\delta=(C_2/C_1-2)/2$
    which indicate the deviation from SPS. Assumed was a Gaussian
    onsite energy disorder.}
  \label{fig:C2_C1}
\end{figure}

We solve Eqs.~\eq{gls_xi_chi_1} and \eq{gls_xi_chi_2} for the ratio of
the cumulants:
\begin{align}
  \label{eq:C2_C1}
  \frac{C_2}{C_1}=\frac{2\chi^{-1}-4\xi^{-1}}{4\xi^{-1}-\chi^{-1}}.
\end{align}
We can compare this with the relation predicted by SPS theory,
Eq.~\eq{C2_C1_SPS}, by examining the parameter
\begin{align}
\label{eq:def_delta}
\delta = \frac{\left|\frac{C_2}{C_1}-2\right|}{2},
\end{align}
which measures the relative deviation from SPS. Fig.~\ref{fig:C2_C1} displays
contour lines $\delta(\sigma,E)=\text{const.}$ for Gaussian onsite disorder, i.e.\ $\nu^3=0,\kappa^4=3\sigma^4$.

The interpretation of Fig.~\ref{fig:C2_C1} is the following. For
disorder strengths $\sigma$ below the contour lines $\sigma<\sigma_\delta(E)$, the relative
deviation from the SPS hypothesis is less than $\delta$. E.g.\ all
points $(\sigma,E)$ below the black line in Fig.~\ref{fig:C2_C1}
fulfill $\delta<10^{-4}$.

\emph{For weak disorder and energies $0<|E|<2$}, we find
$\chi^{-1}=3\xi^{-1}$ which according to Eq.~\eq{C2_C1}
 leads to SPS ($C_2/C_1=2$). Using Eq.~\eq{lambda_final}, we get
\begin{align}
\label{eq:lambda_band}
  \lambda^{-1} &=\frac{\sigma^2}{8-2E^2}, \quad \text{for $0<|E|<2$ and $\sigma\rightarrow 0$,}
\end{align}
in agreement with the well-known formula by Thouless\cite{thouless79}
which, in this energy range, is correct also for the discrete Anderson model.\cite{kappus81}

\emph{At the band center $E=0$} we find that SPS is violated even in
the limit $\sigma\rightarrow 0$. Inserting
\begin{align}
  \xi^{-1}=\ln\left[\frac{\sigma^2}{2} + \sqrt{1 + \left(\frac{\sigma^2}{2}\right)^2} \right],\quad \text{for $E=0$}
\end{align}
as well as, for Gaussian onsite energy disorder $\kappa^4=3\sigma^4$,
\begin{align}
\label{eq:chi_E_0}
\chi^{-1}=\ln\left[1 +
       \frac{3}{2}\sigma^4+\sigma^2\sqrt{3 +
         \left(\frac{3}{2}\sigma^2\right)^2}\right],\quad \text{for $E=0$}
\end{align}
into Eq.~\eq{lambda_final} leads to the localization length
\begin{align}
  \label{eq:lambda_band_center}
  \lambda^{-1} = \frac{2-\sqrt{3}}{4}\sigma^2, \quad \text{for $E=0$
    and $\sigma\rightarrow 0$.}
\end{align}
This equation shows the correct power-law behavior
$\lambda^{-1}\sim\sigma^2$ but deviates from the correct
prefactor.\cite{kappus81,derrida84} This suggests that at the band
center the assumption A breaks down and the simple approximation
\eq{lambda_final} between the localization length and the GLEs
$\xi^{-1}$ and $\chi^{-1}$ is no longer valid. This is a signature of
the band center
anomaly.\cite{kappus81,derrida84,schomerus03,deych03,kravtsov10}

\emph{Similarly, at the band edges $|E|=2$}, SPS is violated also for
$\sigma\rightarrow 0$. Eq.~\eq{lambda_final} leads to
\begin{align}
  \label{eq:lambda_band_edge}
\lambda^{-1}=(2^{1/3}-\frac{1}{4}42^{1/3})
 \times\sigma^{2/3}, \quad \text{for $|E|=2$ and
   $\sigma\rightarrow 0$,}  
\end{align}
again showing the correct power-law behavior $\lambda^{-1}\sim \sigma^{2/3}$
but deviating from the correct prefactor for the same reason as above.

\emph{Outside the band $|E|>2$}, examining $\delta$ does not lead to
agreement with SPS for any disorder strength.

In other words, if assumption A is applied as an approximation to the
discrete Anderson model one obtains the correct localization length in
leading order of $\sigma$ as
long as $E$ is in between the band edges and the band center. Then,
assumption A automatically leads to fulfillment of assumption B.

\section{Conclusions}
\label{sec:conclusions}

The present article deals with the discrete-space 1D Anderson model with
uncorrelated diagonal disorder, characterized by its finite second through
fourth moments $\sigma^2$, $\nu^3$ and $\kappa^4$.

When subjected to a statistical model for the effects of decoherence
(parameters $p=1/L_\phi$, the inverse coherence length, and $\eta$, the energy
broadening due to decoherence),\cite{zilly09}
the length-dependent resistance at infinitesimal bias voltage
undergoes a decoherence-induced transition from localized,
exponentially increasing to ohmic, linearly increasing behavior,
i.e.~we find decoherence-induced conductivity. 

For the ohmic regime we derive the exact value of the resistivity
$\rho$ which for given model parameters $p$ and $\eta$ only depends on
the electron energy $E$ and the second moment $\sigma^2$ of the onsite energy disorder. Higher
moments do not enter.

The critical decoherence density $p^\ast$, at which the transition
from localized to ohmic behavior takes place, is connected to the
second-order generalized Lyapunov exponent (GLE)
$\xi^{-1}=-\ln(1-p^\ast)$.

Therefore we can state the condition for
finite conductivity as $L_\phi<(1-\exp(-\xi^{-1}))^{-1}$, which for weak
disorder and energies inside the band becomes $L_\phi<\xi$, i.e.\
that the phase coherence length be smaller than the length $\xi$. In this way, a property of the
completely coherent Anderson model ($\xi$) determines the limits
($p^\ast$) of the ohmic behavior in the Anderson model including
decoherence.

The GLE $\xi^{-1}$ describes the exponential increase of
$\langle 1/T_j \rangle$, the disorder-averaged dimensionless
resistance. Using the generating function of the
$\langle 1/T_j\rangle$ (which is nothing but $\rho/p^2$) we derive the exact value of $\xi^{-1}$ for
arbitrary electron energy and disorder.

Analogously to $\xi^{-1}$, the fourth-order GLE
$\chi^{-1}$ is calculated from the generating function of $\langle
1/T_j^2\rangle$. In a similar way, all higher even-order GLEs can be determined.

Under the assumption of a Gaussian limiting distribution for $\ln T$
(assumption A), $\xi$ and $\chi$ together allow to determine the
localization length $\lambda$ and the ratio of the first two cumulants
of $\ln T$, $C_2/C_1$ approximately. By studying this ratio we
conclude that the single parameter scaling (SPS) hypothesis which
states $C_2/C_1=2$ is automatically fulfilled in weak disorder for
energies in between the band center and the band edges. Furthermore in
this energy range assumption A leads to the correct weak disorder
limit of the localization length. For this energy range, we present a
maximum disorder $\sigma_\delta(E)$ which yields agreement with SPS,
given a tolerance $\delta$ of deviation.

For the energies $E=0$ and $|E|=2$ (band edges and band center of the
pure system), even for weak disorder the SPS hypothesis is not
fulfilled. Furthermore, the localization length one obtains under
assumption A does not agree with the literature
results.\cite{kappus81,derrida84} We
conclude that at these energies assumption A breaks down, in accordance with Ref.\ \onlinecite{schomerus03}.

\begin{acknowledgments}
  This work was supported by Deutsche Forschungsgemeinschaft under
  grants Nos.~GRK 1240 ``nanotronics'' and SPP 1386 ``nanostructured
  thermoelectric materials''.  O.U.\ acknowledges the support of the
  Alexander von Humboldt Foundation and the J\'anos Bolyai Research
  Foundation of the Hungarian Academy of Sciences. M.W.\ wishes to
  thank Ludger Santen for the support by Saarland University.
\end{acknowledgments}

\appendix

\section{Derivation of the recursive formula for the Green function
  element {\boldmath $G_{1j}$}}
\label{sec:Green_iter}

In this section we derive Eqs.\ \eq{1_T_j_r} and \eq{iteration_r_j}
which are necessary for the calculation of the transmission
function. The Green function of the finite chain with wide-band
(virtual) contacts is the inverse of a tridiagonal, symmetric matrix:
\begin{align}
  \label{eq:Green_inverse_tridiagonal}
  G^{-1} = \mat{ccccc}{ E -\epsilon_1 + \iu \eta
    & -1 & 0 & \cdots & 0 \\ -1 & E-\epsilon_2 & -1  & 0 & \vdots \\ 0 & -1
    & \ddots &  & \\ \vdots &&&& -1 \\ 0 &
    \cdots& 0&  -1 & E - \epsilon_j + \iu \eta}
\end{align}
As such, its components are easily derived by LU decomposition, 
\begin{align}
  \label{eq:LU-decomposition}
  G^{-1}&= \mat{cccc}{\alpha_1 & -1 & 0&\dots \\-1 & \ddots & & \\0 &
    &\ddots &-1 \\ \vdots & & -1 & \alpha_j} \\ 
  &= \underbrace{\mat{cccc}{1 & 0& \dots &\\l_1 & 1 &\ddots & \\0 & \ddots&\ddots
    & 0 \\ \vdots &0 & l_{j-1} & 1}}_{L}  
  \underbrace{\mat{cccc}{m_1 & -1&0 & \dots \\0 & m_2 &\ddots & \\\vdots & \ddots &
    \ddots& -1 \\  & \dots & 0 & m_j}}_{U},
\end{align}
where $\alpha_i = E - \epsilon_i + \iu \delta_{i1} \eta + \iu
\delta_{ij} \eta$ for $i=1,\dots,j$, $l_i = -\tfrac{1}{m_i}$ for
$i=1,\dots, j-1$, $m_1=\alpha_1$ and $m_{i}=\alpha_{i}+l_{i-1}$ for
$i=2,\dots,j$.

From 
\begin{align}
  G^{-1} G = LU G = 1
\end{align}
follows that $G_{1j}$ is the first element of the vector $x$ that fulfills
\begin{align}
  L\underbrace{U x}_y = \delta_{ij}.
\end{align}
Forward substitution yields
\begin{align}
  y=\delta_{ij},
\end{align}
and subsequent backward substitution in
\begin{align}
  Ux = \delta_{ij}
\end{align}
gives
\begin{align}
  G_{1j}=x_1=\frac{1}{m_1\times m_2 \times \dots \times m_j}.
\end{align}
Inserting the definitions of the $m_j$ we find the recursive formula
\begin{align}
  G_{1j} = \frac{1}{P_j}
\end{align}
with the polynomials
\begin{align}
\begin{split}
  P_0&=1 \\
  P_1&=E-\epsilon_1 +\iu \eta \\
P_{i} &= (E-\epsilon_i) P_{i-1} - P_{i-2}, \quad i = 2,\dots, j-1 \\
P_j &= (E-\epsilon_j + \iu \eta) P_{j-1}- P_{j-2}.
\end{split}
\end{align}
As $E$, the $\epsilon_i$, and $\eta$ are real quantities, 
\begin{align}
  P_j = r_j -\eta^2 s_{j-1} + \iu \eta (s_j+r_{j-1})
\end{align}
is the subdivision of the polynomial $P_j$ into its real and imaginary
parts, using the iterative polynomials $r_i$ and $s_i$ defined in \eq{iteration_r_j}. With the definition
of the transmission function (Eq.\ \eq{Transmission}) follows Eq.\ \eq{1_T_j_r}.

\section{Derivation of the generating function {\boldmath $\mc{T}_{-2}$ }}
\label{sec:appendix}

Using the relation \eq{1_T_j_r} of the transmission function to the
recursively defined polynomials $r_j$ and $s_j$ \eq{iteration_r_j} we
derive a recursive expression for $\langle 1/T_j^2\rangle$:
\begin{align}
  \label{eq:def_1_T2}
\begin{split}
\left\langle\! \frac{1}{T_j^2} \!\right\rangle &= \frac{1}{16 \eta^4}
\int\limits_{-\infty}^{\infty}\!\!\! \diffd \epsilon_1 \dots \diffd
\epsilon_j \left[\prod\limits_{i=1}^j w(\epsilon_i)\right] \\
& \times \left( r_j^2
  +\eta^2r_{j-1}^2+\eta^2 s_j^2+ \eta^4 s_{j-1}^2 + 2 \eta^2 \right)^2
\end{split}\\
\label{eq:1_T2_rekursbeg}
\begin{split} &=\frac{1}{4} + \frac{1}{16}\left( \eta^{-4} R_j^4 + 2
  R_{j-1}^4 + \eta^4 R_{j-2}^4 \right) \\ &+ \frac{1}{8} \left(
  \eta^{-2} R_{j}^{22} + \eta^2 R_{j-1}^{22} \right)\\
&+\frac{1}{4} \left( \eta^{-2} R_j + 2 R_{j-1} + \eta^2
  R_{j-2} \right) \\ &+ \frac{1}{8} \left( \eta^{-2} S_{jj}^{22}
  + \eta^2 S_{j-1j-1}^{22} \right)\\ & + \frac{1}{8} \left(
  S_{jj-1}^{22} + S_{j-1j}^{22}\right)
\end{split}
\end{align}
where the $R_j$ are defined in \eq{defR_j}, and 
\begin{align}
  \label{eq:defR_j^4} \begin{split}
R_j^4 &\equiv \mb{I}_j[ r_j^4],\\
R_{j}^{22}&\equiv \mb{I}_j[ r_j^2
r_{j-1}^2]\\ &= \mb{I}_{j+1}[s_{j+1}^2 s_{j}^2], \\
S_{jk}^{22}&\equiv\mb{I}_{\max\{j,k\}}[ r_j^2 s_k^2] \end{split}
\end{align}
using the short-hand notation for the integration 
\begin{align}
\mb{I}_j[\cdot] \equiv \int\limits_{-\infty}^{\infty}\!\!\! \diffd \epsilon_1 \dots \diffd
\epsilon_j \left[\prod\limits_{i=1}^j w(\epsilon_i)\right] [\cdot].
\end{align}
Analogously we define
\begin{align}
\begin{split}
  R^{31}_{j} &\equiv \mb{I}_j[ r^3_j r_{j-1}] \\
R^{13}_{j} &\equiv \mb{I}_j[ r_jr^3_{j-1}]  \end{split}
\end{align}
and
\begin{align}
\begin{split}
  S^{211}_{jk} &\equiv \mb{I}_{\max\{j,k\}}[ r_j^2 s_k s_{k-1}] \\
  S^{112}_{jk} &\equiv \mb{I}_{\max\{j,k\}}[ r_j r_{j-1} s_k^2 ]\\
  S^{1111}_{j} &\equiv \mb{I}_j [ r_j r_{j-1} s_j s_{j-1}] \end{split}
\end{align}

By \eq{iteration_r_j} the $R$ and $S$ fulfill the following recursions
\begin{align}
  \begin{split}
  R^4_j &= X_4 R_{j-1}^4 - 4 X_3 R^{31}_{j-1} + 6
  X_2 R^{22}_{j-1} \\ & - 4 X_1 R^{13}_{j-1} +
  R^4_{j-2}             \nonumber \end{split} \displaybreak[1]
\\ \begin{split} 
R^{31}_j &= X_3 R^4_{j-1} - 3 X_2 R^{31}_{j-1} + 3
X_1 R^{22}_{j-1} - R^{13}_{j-1} \nonumber \end{split} \displaybreak[1] \\ \begin{split}
R^{22}_j &= X_2 R^4_{j-1} - 2 X_1 R^{31}_{j-1} +
R^{22}_{j-1} \nonumber \end{split} \displaybreak[1] \\ \begin{split}
R^{13} &= X_1 R^4_{j-1} - R^{31}_{j-1} \nonumber \end{split} \displaybreak[1] \\ \begin{split}
S^{22}_{jj} &= X_4 S^{22}_{j-1j-1} + 4 X_2
S^{1111}_{j-1} + S^{22}_{j-2j-2} \\ &- 2 X_3
(S^{211}_{j-1j-1} + S^{112}_{j-1j-1}) \\ & + X_2
(S^{22}_{j-1j-2} + S^{22}_{j-2j-1}) \\ &- 2 X_1 (
S^{112}_{j-1j-2} + S^{211}_{j-2j-1})
\nonumber \end{split} \displaybreak[1] \\ \begin{split}
  S^{22}_{jj-1} &= X_2 S^{22}_{j-1j-1} - 2 X_1
  S^{112}_{j-1j-1} + S^{22}_{j-2j-1}
\nonumber \end{split} \displaybreak[1] \\ \begin{split}
  S^{22}_{j-1j} &= X_2 S^{22}_{j-1j-1} - 2 X_1
  S^{211}_{j-1j-1} + S^{22}_{j-1j-2}
\nonumber \end{split} \displaybreak[1] \\ \begin{split}
S^{211}_{jj} &= X_3 S^{22}_{j-1j-1} - 2 X_2
S^{112}_{j-1j-1} + X_1 S^{22}_{j-2j-1} \\ &- X_2
S^{211}_{j-1j-1} + 2 X_1 S^{1111}_{j-1} - S^{211}_{j-2j-1}  
\nonumber \end{split} \displaybreak[1] \\ \begin{split}
S^{112}_{jj} &= X_3 S^{22}_{j-1j-1} - 2 X_2
S^{211}_{j-1j-1} + X_1 S^{22}_{j-1j-2} \\ &- X_2
S^{112}_{j-1j-1} + 2 X_1 S^{1111}_{j-1} - S^{112}_{j-1j-2}  
\nonumber \end{split} \displaybreak[1] \\ \begin{split}
S^{211}_{j-1j} &= X_1 S^{22}_{j-1j-1} -
S^{211}_{j-1j-1} \nonumber \end{split} \displaybreak[1] \\ \begin{split}
S^{112}_{jj-1} &= X_1 S^{22}_{j-1j-1} -
S^{112}_{j-1j-1} \nonumber \end{split} \displaybreak[1] \\ \begin{split}
S^{1111}_j &= X_2 S^{22}_{j-1j-1} - X_1 (
S^{211}_{j-1j-1}+ S^{112}_{j-1j-1}) + S^{1111}_{j-1}
\end{split}
\end{align}
where we have defined 
\begin{align}
  \label{eq:def_Xi}
  X_n &= \mb{I}_1 [ (E-\epsilon_1)^n ] \\
  &=
  \begin{cases}
    E^4 + 6 E^2 \sigma^2 - 4 E \nu^3 + \kappa^4 & \text{ for $n=4$} \\
    E^3 + 3 E \sigma^2 - \nu^3 & \text{ for $n=3$} \\
E^2 + \sigma^2 & \text{ for $n=2$} \\
E & \text{ for $n=1$}
  \end{cases}
\end{align}
For the recursions, the following initial conditions apply:
\begin{align}
\begin{split}
  R^4_1 &= X_4, \qquad R^4_0 =1  \\
R^4_{-1}&=R^{31}_0=R^{22}_0=R^{13}_0=0 \\
S^{22}_{11} &= X_2, \qquad S^{112}_{11}= X_1, \qquad
S^{22}_{01}=1 \\
S^{22}_{10}&= S^{211}_{11}=
S^{211}_{01}=S^{112}_{10}=S^{1111}_1=0
\end{split}
\end{align}
With this we can determine the generating functions of the $R$
and $S$:
\begin{flalign}
 \begin{split}
   \label{eq:R^4}
  \mc{R}^4(z) &\equiv \sum_{j=1}^\infty R_j^4 z^{j-1} \\
  &=X_4 + z X_4 \mc{R}^4(z) + z + z^2 \mc{R}^4(z) \\ &- 4 z X_3
  \mc{R}^{31}(z)  + 6 z X_2 \mc{R}^{22}(z) - 4z X_1 \mc{R}^{13}(z)    
\end{split}
\displaybreak[1]  \\ \begin{split}
\mc{R}^{31}(z) &\equiv \sum_{j=1}^\infty R_j^{31} z^{j-1} \\
&=X_3 + z X_3 \mc{R}^4(z) - 3 z X_2 \mc{R}^{31}(z) \\ &+ 3z X_1
\mc{R}^{22}(z) - z \mc{R}^{13}(z)     
  \end{split}
  \displaybreak[1] \\ \begin{split}
    \mc{R}^{22}(z) &\equiv \sum_{j=1}^\infty R_j^{22} z^{j-1} \\
    &=X_2 + z X_2 \mc{R}^4(z) - 2z X_1 \mc{R}^{31}(z) + z
    \mc{R}^{22}(z) 
  \end{split}
  \displaybreak[1] \\ \begin{split}
    \label{eq:R^13}
    \mc{R}^{13}(z) &\equiv \sum_{j=1}^\infty R_j^{13} z^{j-1} \\
    &=X_1 + z X_1 \mc{R}^4(z) - z \mc{R}^{31}(z)  
  \end{split}
\end{flalign}
\begin{flalign}
  \begin{split}
    \label{eq:G^22_11}
    \mc{S}^{22}_{11}(z) &\equiv \sum_{j=1}^\infty S_{jj}^{22}
    z^{j-1} \\
    &= X_2 + z X_4 \mc{S}^{22}_{11}(z) + 4z X_2 \mc{S}^{1111}(z)
    \\
    &-2z X_3 ( \mc{S}^{211}_{11}(z) + \mc{S}^{112}_{11}(z) )\\ & + z
    X_2 ( \mc{S}^{22}_{10}(z) + \mc{S}^{22}_{01}(z)) \\ &-2 X_1 (
    \mc{S}^{112}_{10}(z) + \mc{S}^{211}_{01}(z)) + z^2 \mc{S}^{22}_{11}(z)
  \end{split} 
  \displaybreak[1] \\ \begin{split}
    \mc{S}^{22}_{10}(z) &\equiv \sum_{j=1}^\infty S_{jj-1}^{22}
    z^{j-1} \\
    &=z X_2 \mc{S}^{22}_{11}(z) - 2z X_1 \mc{S}^{112}_{11}(z) + z \mc{S}^{22}_{01}(z)
  \end{split}
  \displaybreak[1] \\ \begin{split}
    \mc{S}^{22}_{01}(z) &\equiv \sum_{j=1}^\infty S_{j-1j}^{22}
    z^{j-1} \\
    &=1+ z X_2 \mc{S}^{22}_{11}(z) - 2z X_1 \mc{S}^{211}_{11}(z) + z \mc{S}^{22}_{10}(z)
  \end{split}
  \displaybreak[1] \\ \begin{split}
    \mc{S}^{211}_{11}(z) &\equiv \sum_{j=1}^\infty S_{jj}^{211}
    z^{j-1} \\
    &=z X_3 \mc{S}^{22}_{11}(z) - 2z X_2 \mc{S}^{112}_{11}(z) 
    +z X_1\mc{S}^{22}_{01}(z) \\ &- z X_2 \mc{S}^{211}_{11}(z) + 2 z X_1
    \mc{S}^{1111}(z)- z \mc{S}^{211}_{01}(z)
  \end{split}
  \displaybreak[1] \\ \begin{split}
    \mc{S}^{112}_{11}(z) &\equiv \sum_{j=1}^\infty S_{jj}^{112}
    z^{j-1} \\
    &= X_1 + z X_3 \mc{S}^{22}_{11}(z) - 2z X_2 \mc{S}^{211}_{11}(z)  + z X_1 \mc{S}^{22}_{10}(z) \\
    &- zX_2 \mc{S}^{112}_{11} (z)+ 2z X_1
    \mc{S}^{1111}(z) - z \mc{S}^{112}_{10}(z) 
  \end{split}
  \displaybreak[1] \\ \begin{split}
    \mc{S}^{211}_{01}(z) &\equiv \sum_{j=1}^\infty S_{j-1j}^{211}
    z^{j-1} \\
    &=zX_1 \mc{S}^{22}_{11}(z) - z \mc{S}^{211}_{11}(z)
\end{split}
  \displaybreak[1] \\ \begin{split}
    \mc{S}^{112}_{10}(z) &\equiv \sum_{j=1}^\infty S_{jj-1}^{112}
    z^{j-1} \\
    &=zX_1 \mc{S}^{22}_{11}(z) - z \mc{S}^{112}_{11}(z)
\end{split}
  \displaybreak[1] \\ \begin{split}
    \label{eq:G^1111}
    \mc{S}^{1111}(z) &\equiv \sum_{j=1}^\infty S_{j}^{1111}
    z^{j-1} \\
    &=z X_2 \mc{S}^{22}_{11}(z) - z X_1 (\mc{S}^{211}_{11}(z) +
    \mc{S}^{112}_{11}(z)) \\ &+ z \mc{S}^{1111}(z)
\end{split}
\end{flalign}

Eqs.~\eq{R^4}--\eq{R^13} and \eq{G^22_11}--\eq{G^1111} constitute
linear systems of equations for the generating functions $\mc{R}$ and
$\mc{S}$, respectively. 

The solutions read
\begin{flalign}
  \begin{split}
  \label{eq:R^4_loes}
  \mc{R}^4(z) &= \frac{1}{N_2} \left[ -z^4+ \left( -36 X_1^2X_2+3
      X_2-X_4 \right.\right.\nonumber \end{split} \displaybreak[1] \\ \begin{split} &\left.\left. +8 X_1
X_3+1+6 X_2^2-6 X_1^2+24 X_1^4
 \right) z^3 \right.\nonumber \end{split} \displaybreak[1] \\ \begin{split} &\left.+ \left( -6 X_1^2X_4+3 X_2X_4-8
 X_1X_3+12 X_1^2X_2 \right.\right.\nonumber \end{split}
\displaybreak[1] \\ \begin{split} &\left. \left. -3 X_2-18 X_2^3
+1-4 X_1^2+X_4-4 X_3^2 \right.\right.\nonumber \end{split}
\displaybreak[1] \\ \begin{split} &\left. \left.+24 X_1 X_2 X_3
 \right) z^2+ \left( -1-3 X_2X_4+4 X_1^2 \right.\right.\nonumber \end{split}
\displaybreak[1] \\ \begin{split} &\left. \left. + X_4
-6 X_2^2+4 X_3^2 \right) z-X_4 \right]
\end{split}  
\displaybreak[1] \\ \begin{split}
  \label{eq:R^22_loes}
  \mc{R}^{22}(z) &= \frac{1}{N_2} \left[ \left(
      -2 X_1^2+X_2 \right) z^2+ \left(
                      2 X_1X_3-3 X_2^2 \right) z \right.\nonumber \end{split}
\displaybreak[1] \\ \begin{split} &\left. -X_2 
 \right]
\end{split} 
\intertext{and}
\begin{split}
  \label{eq:G^22_11_loes}
  \mc{S}^{22}_{11}(z) &= \frac{1}{N_2} \left[\left(
      -2 X_1^2+X_2 \right) z^2+ \left(
      2 X_1X_3-3 X_2^2 \right) z \right.\nonumber \end{split}
\displaybreak[1] \\ \begin{split} &\left. -X_2 
 \right]
\end{split} \displaybreak[1] \\ \begin{split}
  \label{eq:G^22_10_loes}
  \mc{S}^{22}_{10}(z)&= \frac{1}{N_1 N_2
    (z-1)} \left[ -z^8+ \left( 24 X_1^4-36 X_1^2X_2
    \right. \right.\nonumber \end{split}
\displaybreak[1] \\ \begin{split} &\left.\left. -4 X_1
^2+8 X_1X_3+6 X_2^2-X_4+2 X_2+1
 \right) z^7 \right.\nonumber \end{split}
\displaybreak[1] \\ \begin{split} &\left.+ \left( -48 X_1^6+96 X_1^4X_2+8
 X_1^4-16 X_1^3X_3     \right. \right.\nonumber \end{split}
\displaybreak[1] \\ \begin{split} &\left.\left. -48 X_1^2X_2^2
-4 X_1^2X_4+4 X_1^2X_2+32 X_1X_2X_3  \right. \right.\nonumber \end{split}
\displaybreak[1] \\ \begin{split} &\left.\left. -12 X_2^3-6 X_1^2-8 X_1X_3-4 X_3
^2+2 X_2X_4     \right. \right.\nonumber \end{split}
\displaybreak[1] \\ \begin{split} &\left.\left.+2 X_2^2+X_4-2 X_2+3 \right) 
z^6 + \left( 32 X_1^6+12 X_1^4X_4     \right. \right.\nonumber \end{split}
\displaybreak[1] \\ \begin{split} &\left.\left.-72 X_1^4X_2-48 X_1^3X_2X_3+36 X_1^2X_2
^3-28 X_1^4     \right. \right.\nonumber \end{split}
\displaybreak[1] \\ \begin{split} &\left.\left. +24 X_1^3X_3+8 X_1^2X_
3^2-12 X_1^2X_2X_4+24 X_1^2X_2^
2     \right. \right.\nonumber \end{split}
\displaybreak[1] \\ \begin{split} &\left.\left.+24 X_1X_2^2X_3-18 X_2^4-2 X_1^2X
_4+46 X_1^2X_2     \right. \right.\nonumber \end{split}
\displaybreak[1] \\ \begin{split} &\left.\left. -12 X_1X_2X_3-4
      X_2X_3^2 +3 X_2^2 X_4+2 X_2^3     \right. \right.\nonumber \end{split}
\displaybreak[1] \\ \begin{split} &\left.\left.+14 X_1^2
-8 X_1X_3+4 X_3^2-2 X_2 X_4-16 X_2^2
    \right. \right.\nonumber \end{split}
\displaybreak[1] \\ \begin{split} &\left.\left. +2 X_4-4 X_2-3 \right)
    z^5+ \left( -8 X_1^4X_4+8 X_1^4X_2     \right. \right.\nonumber \end{split}
\displaybreak[1] \\ \begin{split} &\left.\left. +32 X_1^3X_2X_3-24 X_
1^2X_2^3-20 X_1^4-8 X_1^3X_3     \right. \right.\nonumber \end{split}
\displaybreak[1] \\ \begin{split} &\left.\left.-12 X
_1^2X_3^2+10 X_1^2X_2X_4+16 X_1
^2X_2^2     \right. \right.\nonumber \end{split}
\displaybreak[1] \\ \begin{split} &\left.\left. -4 X_1X_2^2X_3+3 X_2^4+6 
X_1^2X_4-2 X_1^2X_2     \right. \right.\nonumber \end{split}
\displaybreak[1] \\ \begin{split} &\left.\left.-28 X_1X_2X_3+
4 X_2X_3^2-3 X_2^2X_4+14 X_2^3     \right. \right.\nonumber \end{split}
\displaybreak[1] \\ \begin{split} &\left.\left. +8 
X_1^2+8 X_1X_3+4 X_3^2-2 X_2 X_4-X_2^2-2 X_4     \right. \right.\nonumber \end{split}
\displaybreak[1] \\ \begin{split} &\left.\left.+4 X_2-3 \right) z^4 +
    \left( 16 X_1^4+4 X_1^2X_3^2-2 X_1^2X_2 X_4     \right. \right.\nonumber \end{split}
\displaybreak[1] \\ \begin{split} &\left.\left.-8 
X_1^2X_2^2-4 X_1X_2^2X_3+3 X_2
^4+2 X_1^2X_4     \right. \right.\nonumber \end{split}
\displaybreak[1] \\ \begin{split} &\left.\left.-14 X_1^2X_2+4 X_1
X_2X_3-2 X_2^3-12 X_1^2-4 X_3^2     \right. \right.\nonumber \end{split}
\displaybreak[1] \\ \begin{split} &\left.\left.+2 X_2 X_4+11
      X_2^2-X_4+2 X_2+3 \right) z^3     \right.\nonumber \end{split} 
\displaybreak[1] \\ \begin{split} &\left.+ \left( -2 X_1^2X_4+2 X_1^2X_2+4 X_1X_2X_
3-2 X_2^3     \right. \right.\nonumber \end{split}
\displaybreak[1] \\ \begin{split} &\left.\left.-2 X_1^2-X_2^2+X_4-2
 X_2+1 \right) z^2+ \left( 2 X_1^2     \right. \right.\nonumber \end{split}
\displaybreak[1] \\ \begin{split} &\left.\left.-X_2^2-1
 \right) z
 \right]
\end{split}  \displaybreak[1] \\ \begin{split}
  \label{eq:G^22_01_loes}
  \mc{S}^{22}_{01}(z)&= \frac{1}{N_1 N_2
    (z-1)} \left[  -z^7+ \left( 20 X_1^4-32 X_1^2X_2 -4 X_1
^2 \right. \right.\nonumber \end{split}
\displaybreak[1] \\ \begin{split} &\left.\left.+8 X_1X_3+5 X_2^2-X_4+2 X_2+1
 \right) z^6+ \left( -16 X_1^6 \right. \right.\nonumber \end{split}
\displaybreak[1] \\ \begin{split} &\left.\left.+48 X_1^4X_2-8
 X_1^3X_3-36 X_1^2X_2^2-4 X_1^2
X_4 \right. \right.\nonumber \end{split}
\displaybreak[1] \\ \begin{split} &\left.\left.+8 X_1^2X_2+28 X_1X_2 X_3-10 X_2
^3-4 X_1^2 \right. \right.\nonumber \end{split}
\displaybreak[1] \\ \begin{split} &\left.\left.-8 X_1X_3-4 X_3^2+2
      X_2X_4+2 X_2^2+X_4-2 X_2 \right. \right.\nonumber \end{split} 
\displaybreak[1] \\ \begin{split} &\left.\left.+3 \right) z^5+ \left( 4 X_1^4X_4-16 X_1^4X_2-16 X_1^3
X_2X_3 \right. \right.\nonumber \end{split}
\displaybreak[1] \\ \begin{split} &\left.\left.+12 X_1^2X_2^3-16 X_1^4+8 X_1
^3X_3+4 X_1^2X_3^2 \right. \right.\nonumber \end{split}
\displaybreak[1] \\ \begin{split} &\left.\left.-8 X_1^2X_2X_4+16
      X_1^2X_2^2+20 X_1X_2^2X_3-15 
 X_2^4 \right. \right.\nonumber \end{split}
\displaybreak[1] \\ \begin{split} &\left.\left.+32 X_1^2X_2-12 X_1X_2X_3-4 
X_2X_3^2+3 X_2^2 X_4 \right. \right.\nonumber \end{split}
\displaybreak[1] \\ \begin{split} &\left.\left.+2 X_2^3+12 X_
1^2-8 X_1X_3+4 X_3^2-2 X_2X_4 \right. \right.\nonumber \end{split}
\displaybreak[1] \\ \begin{split} &\left.\left.-13 X_
2^2+2 X_4-4 X_2-3 \right) z^4+ \left( -4 X_1^4 \right. \right.\nonumber \end{split}
\displaybreak[1] \\ \begin{split} &\left.\left.-4 X_1^2X_3^2+4
      X_1^2X_2 X_4+4
 X_1^2X_2^2-4 X_1X_2^2 X_3 \right. \right.\nonumber \end{split}
\displaybreak[1] \\ \begin{split} &\left.\left.+3 X_2
^4+4 X_1^2X_4-4 X_1^2X_2-20 X_1X_2X_3 \right. \right.\nonumber \end{split}
\displaybreak[1] \\ \begin{split} &\left.\left.+4 X_2X_3^2-3 X_2^2X_4+10 X_2
^3+4 X_1^2+8 X_1X_3 \right. \right.\nonumber \end{split}
\displaybreak[1] \\ \begin{split} &\left.\left.+4 X_3^2-2 X_2X_4
      -X_2^2-2 X_4+4 X_2-3 \right) z^3 \right.\nonumber \end{split} 
\displaybreak[1] \\ \begin{split} &\left.+ \left( -4 X_1^2X_2+4 X_1X_2X_3-2 X_2^3-8 
X_1^2-4 X_3^2 \right. \right.\nonumber \end{split}
\displaybreak[1] \\ \begin{split} &\left.\left.+2 X_2X_4+8 X_2^2-X_4
+2 X_2+3 \right) z^2+ \left( -X_2^2 \right. \right.\nonumber \end{split}
\displaybreak[1] \\ \begin{split} &\left.\left.+X_4-2 X_
2+1 \right) z-1
 \right]
\end{split} 
\intertext{where}
\begin{split}
  \label{eq:Nenner_G_1}
  N_1 &= z^3+ \left( 1-2 X_1^2+X_2 \right) z^2+ \left( -1+
X_2 \right) z-1 \end{split}  \displaybreak[1] \intertext{was already
presented in Eq.~\eq{def_N_1} and} \begin{split}
  \label{eq:Nenner_R}
  N_2 &=z^5+ \left( -8 X_1 X_3+36 X_1^2X_2-6 X_2^2-24 X_1^4
  \right.\nonumber \end{split} \displaybreak[1] \\ \begin{split} &\left. -3 X_2+6 X_1^2-1+X_4
 \right) z^4+ \left( 6 X_1^2X_4+3 X_2 \right.\nonumber \end{split}
\displaybreak[1] \\ \begin{split} &\left. -3 X_2
X_4-2-X_4+18 X_2^3+8 X_1X_3+4 X_1^2 \right.\nonumber \end{split} \displaybreak[1] \\ \begin{split} &\left.
-24 X_1X_2X_3+4 X_3^2-12 X_1^2X_2
 \right) z^3+ \left( -10 X_1^2 \right.\nonumber \end{split} \displaybreak[1] \\ \begin{split} &\left.+6 X_2^2-X_4+2
-4 X_3^2+3 X_2X_4+3 X_2 \right) z^2 \nonumber \end{split} \displaybreak[1] \\ \begin{split} &+
 \left( X_4-3 X_2+1 \right) z-1.
\end{split} 
\end{flalign}
Here we have only displayed those generating
functions which enter into the generating function of
$\langle 1/T_j^2\rangle $ defined below. The other functions
$\mc{R}(z)$ and $\mc{S}(z)$ have similar expressions, particularly
the same denominators.

Using \eq{1_T2_rekursbeg} and the generating functions for the
$\mc{R}$ and $\mc{S}$ \eq{R^4_loes}--\eq{G^22_01_loes} we arrive at the
generating function of $\langle 1/T_j^2 \rangle $:

\begin{flalign}
  \mc{T}_{-2}(z) &\equiv \sum_{j=1}^\infty
  \left\langle\!\frac{1}{T_j^2}\!\right\rangle  z^{j-1} 
\nonumber \displaybreak[1] \\ \begin{split}
  \label{eq:generating_1_T2}
  &=\frac{5}{8} + \frac{1}{4} \frac{1}{1-z} +
  \left(\frac{\eta^2}{4}+\frac{\eta^4}{16}\right) z  \\
  &+\left(\frac{1}{16\eta^4} + \frac{1}{8}z + \frac{\eta^4}{16}z^2
  \right) \mc{R}^4(z) \\
  &+\left(\frac{1}{8\eta^2} + \frac{\eta^2}{8}z \right)
  \left(\mc{R}^{22}(z)+ \mc{S}^{22}_{11}(z) \right)
  \\
&+\left(\frac{1}{4\eta^2}+ \frac{1}{2}z + \frac{\eta^2}{4}z^2\right)
\mc{R}(z) \\ 
&+\frac{1}{8}\left(\mc{S}^{22}_{10}(z)+\mc{S}^{22}_{01}(z)\right),
\end{split} 
\end{flalign}
where $\mc{R}(z)$ was defined in Eq.~\eq{def_F}.

\section{Calculation of the generating function of {\boldmath $\langle
     1/T_j^2\rangle $ without coupling to external leads}}
\label{sec:without_leads}

Using the Green function of the tight-binding Anderson Hamiltonian
without coupling to the environment,
\begin{align}
  \label{eq:Green_wo_coupling}
  G \equiv \lim\limits_{\eta \rightarrow 0} \left[ E - H + \iu \eta
    \mathbf{1} \right]^{-1},
\end{align}
$\mathbf{1}$ being the unit matrix of adequate size, one defines the
transmission probability of an electron between sites $|1\rangle$ and
$|j\rangle$ as
\begin{align}
  \label{eq:T_wo_coupling}
  T_j \equiv |\langle 1| G | j\rangle|^2.
\end{align}

Similar to Eqs.~\eq{1_T_j_r} and \eq{iteration_r_j} a recursive
calculation of the transmission is possible:
\begin{align}
  \label{eq:1_T_j_wo_coupling}
  \frac{1}{T_j} = \lim\limits_{\eta \rightarrow 0} \left(R_j^2 + I_j^2\right),
\end{align}
where the $R_j$ and $I_j$ fulfill
\begin{align}
  \label{eq:recursion_wo_coupling}
\begin{split}
  R_j &= (E-\epsilon_j) R_{j-1} - \eta I_{j-1} - R_{j-2}, \\
  I_j &= (E-\epsilon_j) I_{j-1} + \eta R_{j-1} - I_{j-2}.
\end{split}
\end{align}
Into the calculation of 
\begin{align}
  \label{eq:F_wo_leads}
  \mc{T}_{-1}(z) \equiv \sum\limits_{j=1}^\infty \left \langle\!
    \frac{1}{T_{j}}\!\right \rangle z^{j-1}
\end{align}
enter the 7 generating functions of the disorder averages of the
two-factor products
$R_j^2$, $R_jR_{j-1}$, $I_j^2$, $I_jI_{j-1}$, $R_jI_j$, $R_jI_{j-1}$, $I_jR_{j-1}$ . The solution reads
\begin{align}
  \label{eq:F_wo_loes}
  \mc{T}_{-1}(z)={\frac {-z^2 + (E^2-\sigma^2-1)z -E^2-\sigma^2}{z^3+ (\sigma^2-E^2+1) z^{2
}+(\sigma^2+E^2-1) z - 1}}
\end{align}
and has the same poles as $\mc{R}$ in Eq.~\eq{def_F}. 

Similarly, into the calculation of 
\begin{align}
  \label{eq:S_wo_leads}
  \mc{T}_{-2}(z) \equiv \sum\limits_{j=1}^\infty \left \langle\!
    \frac{1}{T_{j}^2}\!\right \rangle z^{j-1}
\end{align}
enter the 30 generating functions of the disorder averages of all
four-factor products composed of
$R_j$, $I_j$, $R_{j-1}$, and $I_{j-1}$. 

The closed form of $\mc{T}_{-2}(z)$ is a very long expression, yet its
denominator reads
\begin{align}
  \label{eq:denominator_S_wo_leads}
  N_1^3\times N_2^5 \times(z-1)
\end{align}
with the same $N_1$ and $N_2$ as defined above in Eqs.~\eq{def_N_1}
and \eq{Nenner_R}.

Therefore, the generating functions $\mc{T}_{-1}(z)$ and $\mc{T}_{-2}(z)$ have
the same poles $q^\ast$ and $z^\ast$ as reported above, and we find
the same asymptotic behavior of $\left \langle 1/T_j \right
\rangle$ and $\left \langle 1/T_j^2 \right
\rangle$ as $j\rightarrow \infty$. Hence the generalized Lyapunov
exponents $\xi^{-1}$ and $\chi^{-1}$ of an
infinite system are independent of whether one takes an environment
into account or not.

\bibliographystyle{apsrev}

\end{document}